# Analysis of Seismocardiographic Signals Using Polynomial Chirplet Transform and Smoothed Pseudo Wigner-Ville Distribution

Amirtaha Taebi, *Student Member, IEEE*, Hansen A. Mansy
Biomedical Acoustics Research Laboratory, University of Central Florida, Orlando, FL 32816, USA
{taebi@knights., hansen.mansy@} ucf.edu

*Abstract*—Seismocardiographic (SCG) signals are chest surface vibrations induced by cardiac activity. These signals may offer a method for diagnosing and monitoring heart function. Successful classification of SCG signals in health and disease depends on accurate signal characterization and feature extraction. One approach of determining signal features is to estimate its time-frequency characteristics. In this regard, four different time-frequency distribution (TFD) approaches were used including short-time Fourier transform (STFT), polynomial chirplet transform (PCT), Wigner-Ville distribution (WVD), and smoothed pseudo Wigner-Ville distribution (SPWVD). Synthetic SCG signals with known time-frequency properties were generated and used to evaluate the accuracy of the different TFDs in extracting SCG spectral characteristics. Using different TFDs, the instantaneous frequency (IF) of each synthetic signal was determined and the error (NRMSE) in estimating IF was calculated. STFT had lower NRMSE than WVD for synthetic signals considered. PCT and SPWVD were, however, more accurate IF estimators especially for the signal with time-varying frequencies. PCT and SPWVD also provided better discrimination between signal frequency components. Therefore, the results of this study suggest that PCT and SPWVD would be more reliable methods for estimating IF of SCG signals. Analysis of actual SCG signals showed that these signals had multiple spectral components with slightly time-varying frequencies. More studies are needed to investigate SCG spectral properties for healthy subjects as well as patients with different cardiac conditions.

## I. Introduction

Cardiovascular disease is a leading cause of death in the United States that accounts for 24.2% of total mortality in 2010 [1]. Associated death rate may be decreased by improving current cardiac diagnostic and patient monitoring methods. Therefore, analysis of blood flow dynamics [2], [3] and the heart related signals [4] has become an active area of research. Auscultation of heart sounds is a widely used procedure performed during physical examinations and can provide useful diagnostic information. Computer analysis of heart sound and vibration provides additional quantitative information that may be helpful for detecting a variety of heart conditions. Cardiac vibrations measured non-invasively at the chest surface are called seismocardiographic (SCG) signals. These signals are believed to be caused by mechanical activities of the heart (e.g. valves closure, cardiac contraction, blood momentum changes, etc.). The characteristics of these signals were found to correlate with different cardiovascular pathologies [4]–[7]. These signals may also contain useful information that is complementary to other diagnostic methods [8]–[10]. For example, due to the mechanical nature of SCG, it may contain information that is not manifested in electrocardiographic (ECG) signals. A typical SCG contains two main events during each heart cycle. These events can be called the first SCG (SCG1) and the second SCG (SCG2). These SCG events contain relatively low-frequency components. Since the human auditory sensitivity decreases at low frequencies, examination of SCG signals may not be optimally done by the unaided human ear [11], [12]. Therefore, a computer assisted analysis may help find possible correlations between the SCG signal characteristics and cardiovascular conditions. Many methods have been used to investigate the SCG features [13]–[15] including time-frequency distributions (TFD) [16]. The current study aims at comparing the performance of some available TFD methods for estimating the spectral content of SCG signals.

TFDs have been used to identify the time-frequency features of a wide range of biomedical signals such as ECG [17], EEG [18], PCG [19], and SCG [16]. The performance of each TFD technique depends on its underlying assumptions, which can consequently lead to different levels of accuracy in estimating spectral content of the signals under consideration. One common approach for TFD estimation is the short-time Fourier transform (STFT). STFT is relatively simple but can't track steep signal temporal changes [19], [20]. In addition, the Heisenberg uncertainty principle leads to resolution limitations. For instance, improving the STFT resolution in time domain worsens the resolution in the frequency domain and vice versa.

The Wigner-Ville distribution (WVD) is another TFD method that was proposed for time-frequency analysis of non-stationary signals [21]. The WVD provides a finer resolution in both time and frequency directions compared to STFT at the cost of the introduction of artifact peaks and increased aliasing in the time-frequency plane [22], [23]. For example, aliasing in the WVD will take place for frequencies above ¼ of the sampling frequency compared to the ½ of the sampling frequency in STFT [24]. The resulting artifacts may overlap with the actual frequency components, and therefore they may result in a misleading interpretation of the TFD. The artifacts can be reduced by smoothing WVD in both time and frequency. When WVD is only smoothed in the frequency direction, the new distribution is called pseudo WVD and when it is smoothed in both time and frequency, it is known as smoothed pseudo WVD (SPWVD). However, smoothing can negatively affect resolution and a compromise needs to be reached between reduction of interference terms and refining the temporal-spectral resolution. The general properties of the STFT and WVD have been discussed in literature [19], [25], [26].

Chirplet transform (CT) that might be considered a generalization of STFT and wavelet transform (WT) [27], involves a complex function of time, frequency, scale and chirp rate; where the chirp rate can be defined as the instantaneous rate of change in the signal frequency [28]. Since the conventional CT is based on a kernel with linear instantaneous frequency (IF), it may provide inaccurate TFD estimations for signals that have nonlinear IF trajectory. Polynomial chirplet transform (PCT) was proposed to solve this problem [20]. It was developed based on a polynomial nonlinear kernel that makes it more appropriate for the analysis of signals with either linear or nonlinear continuous IF trajectories.

Some previous studies compared the performance of different TFD methods for the analysis of cardiac signals. For instance, Obaidat [19] utilized STFT, WVD, and WT for the analysis of phonocardiogram signals. Comparing the resolution among the different methods of interest, he concluded that WT performed better than STFT and WVD in providing more details of the heart sounds. White [29] exploited the pseudo Wigner-Ville distribution to detect, analyze and classify heart murmurs in PCG signals. Taebi & Mansy [16] employed STFT, PCT and continuous WT with different mother functions to determine the most suitable TFD technique for analysis of SCG signals. They concluded that PCT estimated the instantaneous frequency of the SCG signal with higher accuracy than the other methods considered.

Understanding different characteristics of SCG, including its TFD, may lead to a better understanding of heart function. Furthermore, successful classification of SCG signals in health and diseased can provide a possible method for diagnosing and monitoring cardiac mechanical activities. This paper is aimed at comparing the performance of different methods of estimating the TFD of SCG signals. Here, estimations from two methods (i.e., STFT and PCT) that were found accurate in an earlier study [16] were compared with two additional methods (WVD and SPWVD). Results of this study will help guide the choice of optimal TFD analysis methods of SCG signals. The theory of different TFD techniques and definition of the synthetic test signals used are described in section II. Results are presented and discussed in section III, followed by conclusions in section IV.

## II. METHODOLOGY

The TFD of the signals under consideration was estimated using four different methods: short-time Fourier transform, polynomial chirplet transform, Wigner-Ville distribution, and smoothed pseudo Wigner-Ville Distribution. This section provides the definitions and properties of the TFD techniques of interest as well as descriptions of the synthetic test signals used and the methods of SCG data acquisition.

### A. TFD Methods

STFT: The STFT of a signal $x(t)$ can be expressed as:

$$\bar{X}_{STFT} = \int_{-\infty}^{+\infty} x(\tau) w(\tau - t) e^{-j\omega\tau} d\tau \quad (1)$$

where $j, w(t), t$ and $\omega$ are $\sqrt{-1}$, the window function, time and frequency, respectively. In this TFD method, the signal $x(t)$ is divided into a number of sub-records that are shorter than $x(t)$. The purpose of using the window function, $w$, is to decrease spectral leakage when the Fast Fourier Transform is applied to each sub-record. This approach assumes that the signal in each sub-record is stationary, i.e. the signal is assumed to have non-varying spectral characteristics in each sub-record [30]. When steep signal non-stationarity is absent in all sub-records, high quality TFD estimates are expected for the whole signal duration. When steep non-stationarity is present, the sub-records need to be shortened to reduce the non-stationarity in individual sub-records, which would enhance temporal resolution. This will, however, cause deterioration in the frequency resolution. Hence, refining temporal and frequency resolutions are two competing effects and a compromise will need to be reached to accurately estimate the instantaneous frequency.

WVD: For a signal $x(t)$, with the analytic associate $z(t)$, the WVD is defined as,

$$\bar{X}_{WVD} = \int_{-\infty}^{+\infty} z(t + \frac{\tau}{2}) z^*(t - \frac{\tau}{2}) e^{-j\omega\tau} d\tau \quad (2)$$

where the superscript * denotes the complex conjugate, and t and $\omega$ are time and frequency, respectively. The analytic associate of the signal $x(t)$ is defined as: $z(t) \equiv x(t) + jH[x(t)]$, where $H[x(t)]$ is the Hilbert transform of the signal $x(t)$, and defined as,

$$H[x(t)] = p.v. \int_{-\infty}^{+\infty} x(\tau) h(t - \tau) d\tau = \frac{1}{\pi} p.v. \int_{-\infty}^{+\infty} \frac{x(\tau)}{t-\tau} d\tau \quad (3)$$

where $p.v.$ denotes the Cauchy principal value. Since the WVD separates the signals in both time and frequency directions, it was suggested as a possible method of analysis for non-stationary signals [31], [32].

PCT: For signals where IF is a nonlinear function of time, CT will have limited ability in tracking IF [33]. Therefore, a new CT-based TFD technique with nonlinear frequency rotation and shift operators and a polynomial kernel was proposed to effectively estimate the nonlinear IF trajectory of non-stationary signals [20]. This technique is called PCT and can provide a TFD with finer resolution compared to conventional CT for signals with either linear or nonlinear IF trajectory. More detailed definition of PCT is provided in [20].

### B. Test Signals

Two synthetic test signals that are similar to SCG were generated and analyzed using the TFD techniques under consideration. The test signals were generated such that they have durations and frequency content similar to the actual SCG signals. The synthetic signals had known IF and hence can serve as the gold standard for testing the accuracy of the TFD under consideration. The analysis was aimed at comparing the accuracy of different TFD techniques for estimating the frequency contents of SCG signals. The amplitude, instantaneous frequency and mathematical description of the test signals are listed in Table I.

TABLE I. LIST OF AMPLITUDE, INSTANTANEOUS FREQUENCY AND EQUATION OF THE SIMULATED SIGNALS USED IN THE CURRENT STUDY.

| Signal | Amplitude | Instantaneous Frequency (Hz) | Signal Equation |
|---|---|---|---|
| $x_1$ | $A_1 = \begin{cases} 0 & 0 < t \leq 0.25 \\ 0.5 - 0.5\cos(14\pi(t-0.75)) & 0.25 < t \leq 0.40 \\ 0 & 0.40 < t \leq 0.70 \\ 0.45 - 0.45\cos(14\pi(t-0.75)) & 0.70 < t \leq 0.83 \\ 0 & 0.83 < t \leq 1.00 \end{cases}$ | $IF_1 = 20$ and $40$ | $x_1 = -A_1 \sin(2\pi(20)t + 94) + 0.9 A_1 \sin(2\pi(40)t + 188)$ |
| $x_2$ | $A_2 = \begin{cases} 0 & 0 < t \leq 0.25 \\ 0.5 - 0.5\cos(14\pi(t-0.75)) & 0.25 < t \leq 0.40 \\ 0 & 0.40 < t \leq 0.70 \\ 0.25 - 0.25\cos(14\pi(t-0.75)) & 0.70 < t \leq 0.83 \\ 0 & 0.83 < t \leq 1.00 \end{cases}$ | $IF_2 = 40$ and $2610\,t^2 - 430\,t + 20$ (where $0 \leq t \leq 0.15$) | $x_2 = y_1 + y_2$ where: $y_1 = -0.5 A_2 \sin(2\pi(40)t)$ $y_2 = \begin{cases} 0 & 0 < t \leq 0.25 \\ z & 0.25 < t \leq 0.40 \\ 0 & 0.40 < t \leq 0.70 \\ z & 0.70 < t \leq 0.83 \\ 0 & 0.83 < t \leq 1.00 \end{cases}$ where: $z = A_2 \sin(2\pi(870\,t^2 - 215\,t + 20)t)$ where $0 \leq t \leq 0.15$ |

*C. IF Error Analysis*

The IF equations of synthetic signals (i.e., those with known IFs) were estimated using different TFD techniques. To evaluate the accuracy of the different TFD methods, the root-mean-square error (RMSE) between actual and estimated IF values was calculated as:

$$RMSE = \sqrt{\frac{\sum_{i=1}^{n}(IF_{actual,i} - IF_{estimated,i})^2}{n}} \quad (4)$$

where $IF_{actual,i}$ and $IF_{estimated,i}$ are the signal actual and estimated IF at time *i*, respectively, and *n* is the total number of data points. RMSE values are then normalized by dividing RMSE by the mean actual instantaneous frequency, $\overline{IF}_{actual}$, of each signal as follows:

$$NRMSE = \frac{RMSE}{\overline{IF}_{actual}} \quad (5)$$

Normalized root-mean-square error (NRMSE) was used in the current study as the criterion to quantify the accuracy of the different TFD techniques in estimating IF. Here, lower NRMSE values would indicate higher accuracy. The signal processing steps for evaluating the most appropriate TFD methods for time-frequency analysis of SCG signals are shown in Fig. 1.

*D. Data Acquisition of SCG*

Actual SCG signals from 8 healthy subjects were also analyzed using the TFD techniques under consideration. After IRB approval, a light-weight (2 gm) accelerometer (PCB Piezotronics, Depew, NY) was used to measure the SCG signal. The sensor was placed over the chest of volunteers at the left sternal border and the 4th intercostal space. The signal was digitized at a sampling frequency of 3200 Hz and down-sampled to 320 Hz. Matlab (R2015b, The MathWorks, Inc, Natick, MA) was used to both acquire and process all signals.

## III. RESULTS AND DISCUSSIONS

The time series and TFD of the synthetic test signals under consideration are shown in Fig. 2 and Fig. 3. The TFDs were estimated using STFT, PCT, WVD, and SPWVD and shown in subfigures b, c, d, and e, respectively. The PSD was also calculated from the TFDs, and normalized with respect to the signal energy. PSDs are presented in the left side of the Fig. 2 and Fig. 3. Here, the signals did not have significant energy above 70 Hz. Therefore, the spectral information is shown for frequencies up to 70 Hz. Table II lists the temporal and spectral resolutions of different TFD techniques. The NRMSE between the actual and calculated instantaneous frequency are reported in Table III for different TFDs.

Temporal and frequency resolution: In general, coarser resolution is not desirable as it may lead to higher errors in estimating IF. Table II shows that STFT had coarser temporal resolution compared to other methods (the latter three methods had the same temporal resolution). SPWVD and STFT techniques had the finest and coarsest spectral resolutions, respectively (Table II). The WVD had a finer resolution than PCT (about 0.11 and 0.22 Hz, respectively) at the cost of increased artifacts that may result from cross-terms and aliasing. As expected, aliasing was seen in the current study for the WVD and SPWVD TFD of all signals with folding frequencies of $f_{Nyq.}/2$, where $f_{Nyq.}$ is the Nyquist frequency.

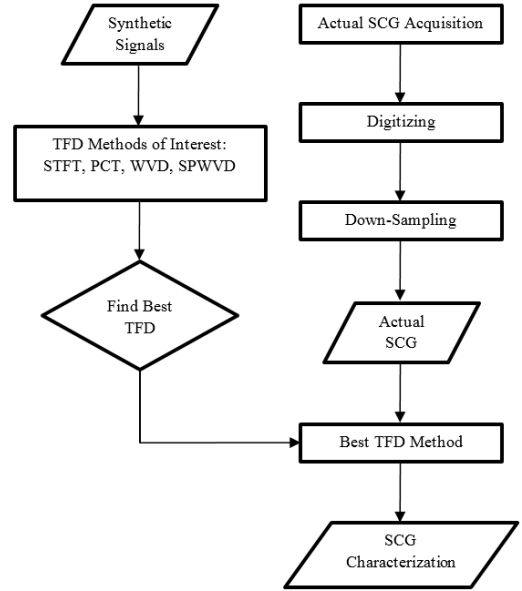

Fig. 1. Block diagram describing the signal processing steps for evaluating the most appropriate TFD methods for time-frequency analysis of the SCG signals.

TABLE II. TEMPORAL AND SPECTRAL RESOLUTION FOR DIFFERENT SIGNALS AND TFD TECHNIQUES FOR FREQUENCIES UP TO 70 HZ. STFT TENDED TO HAVE COARSER TEMPORAL AND SPECTRAL RESOLUTION COMPARED TO PCT. SPWVD HAD THE FINEST TEMPORAL AND SPECTRAL RESOLUTION.

|  |  | STFT | PCT | WVD | SPWVD |
|---|---|---|---|---|---|
| Temporal Resolution (ms) |  | 12.5 | 3.1 | 3.1 | 3.1 |
| Spectral Resolution (Hz) | $x_1$ | 0.6250 | 0.2462 | 0.1231 | 0.0004 |
|  | $x_2$ | 2.5000 | 0.2462 | 0.1231 | 0.0038 |

TABLE III. NRMSE BETWEEN THE ACTUAL AND CALCULATED IF FOR DIFFERENT TFD METHODS. THE MOST APPROPRIATE TFD TECHNIQUES WOULD BE THOSE WITH THE LOWEST NRMSE VALUES.

|  | STFT | PCT | WVD | SPWVD |
|---|---|---|---|---|
| $x_1$ | 0.0199 | 0.0214 | 0.6970 | 0.0325 |
| $x_2$ | 0.4431 | 0.2813 | 1.2202 | 0.2672 |

*A. Synthetic SCG with Time-independent Frequencies*

This synthetic SCG consisted of two frequency components at 20 and 40 Hz. Fig. 2.d shows that WVD TFD was contaminated by relatively strong artifacts between the actual signal frequency components. In addition, WVD did not distinguish between the two components and had noticeable leakage, which led to a PSD with 3 peaks at 20, 30, and 40 Hz. The 30 Hz peak is clearly due additional interference terms. The PSD graphs of STFT, PCT, and SPWVD correctly showed two peaks corresponding to the actual signal components. While STFT and PCT demonstrated less leakage than WVD, they showed possible more leakage than SPWVD where the spectral peaks appeared more concentrated. The WVD had the highest error among the methods (NRMSE of 0.6970), which may be, at least in part, due to the interference terms in WVD. STFT and PCT had lower NRMSE values compared to SPWVD (Table III); however, the latter had the finest spectral resolution and a relatively smaller leakage.

*B. Synthetic SCG with Varying Frequency Components*

This synthetic SCG consisted of two frequency components. The first component had a varying frequency ranging from 20 to 7 Hz while the second component was a fixed frequency at 40 Hz. In addition, the signal was contaminated with white noise with a signal-to-noise ratio of 10. PCT and SPWVD estimated the signal IF with higher accuracy than other methods (NRMSE of 0.2813 and 0.2672, respectively). STFT had a higher NRMSE of 0.4431, and WVD had the lowest accuracy with an NRMSE of 1.2202. WVD had finer temporal and spectral resolutions than PCT and STFT at the cost of introduction of artifact peaks in the time-frequency plane (Fig. 3.d). These artifacts were significantly reduced after smoothing in time and frequency domains (Fig. 3.e). In addition, WVD appeared to have the highest artifacts. This leakage in WVD was also reduced by employing SPWVD. It can also be seen in Fig. 3 that STFT and PCT could distinguish between the two components of the signal; however, they appeared to have slightly higher leakage between the two frequency peaks compared to the SPWVD.

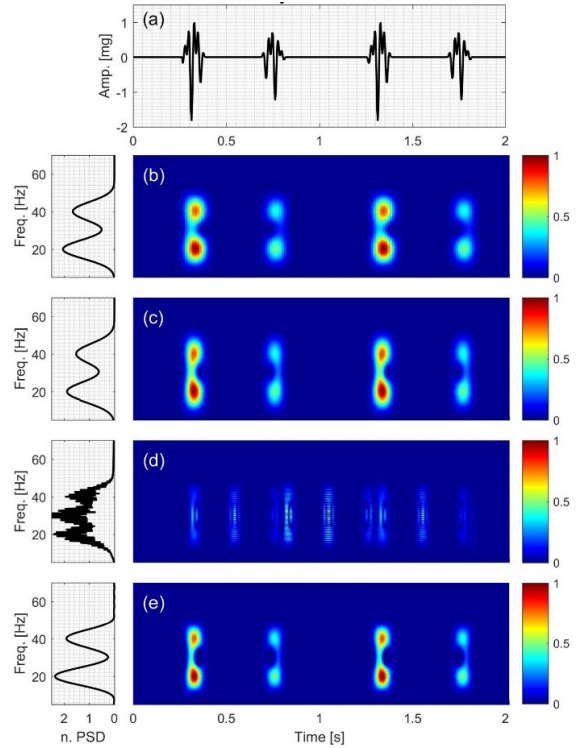

Fig. 2. Synthetic SCG with independent frequencies: (a) Time series. Time-frequency distribution using (b) STFT, (c) PCT, (d) WVD, and (e) SPWVD, respectively.

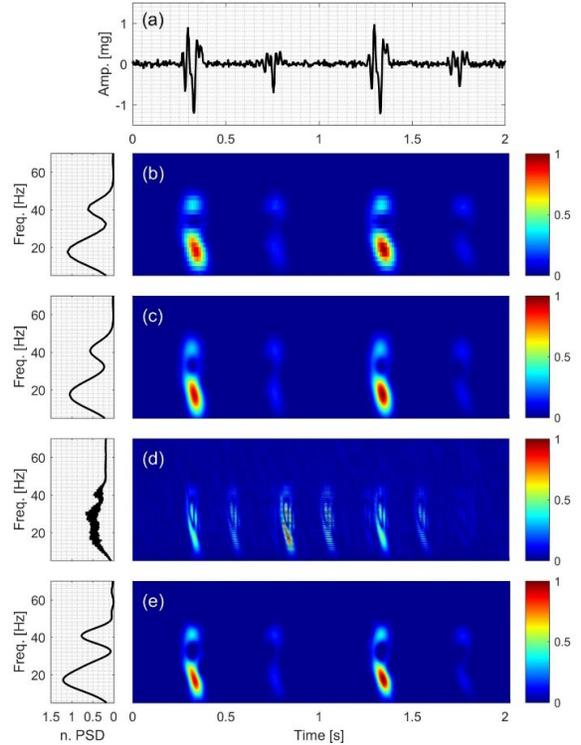

Fig. 3. Synthetic SCG with varying frequency components: (a) Time series. Time-frequency distribution using (b) STFT, (c) PCT, (d) WVD, and (e) SPWVD, respectively.

The STFT is relatively simple and mostly suitable for stationary signals [30]. However, SCG signals are usually rhythmic, and may have different amplitude, time duration, and spectral properties for each cardiac cycle [34], [35]. Thus, STFT may not be the best option for time-frequency analysis of SCG signals. TFD of the synthetic signals in the current study suggest that PCT and SPWVD consistently had low NRMSE, less artifacts and lower leakage. These two techniques also provided better discrimination between signal frequency components. The PCT and SPWVD performance was better than STFT for the synthetic signals, which may be due to their finer resolution. SPWVD had more accurate IF estimations than WVD for the synthetic SCG signals. The performance of STFT varied depending on the synthetic signal under consideration, but it did not seem to provide noticeable advantages over SPWVD or PCT, except for simplicity. These trends will be helpful in the interpretation of the results of the actual SCGs TFD. Since PCT and SPWVD provided more accurate IF values than STFT, we decided to utilize PCT and SPWVD for estimating the frequency content of the actual SCG signals.

*C. Actual SCG Signals*

Fig. 4 shows the TFDs of an actual SCG for two cardiac cycles. The figure suggests that there were two SCG events (SCG1 and SCG2) for each cardiac cycle. Using PCT and SPWVD, SCG1 and SCG2 appeared to be localized in the time-frequency domain at frequencies of 18.75 and 37.50 Hz. This was also clearly seen in the PSD of the PCT and SPWVD (Fig. 4.c and 4.e, respectively). The PSD of the WVD, instead, showed a noisy broadband rather than two separate peaks, which may be due to presence of the artifact peaks. Comparing Fig 4.e with Fig. 2.e and 3.e, one could also conclude that the high-frequency component of the SCG1 behaved more as a fixed frequency, while the lower-frequency component behaved more as a chirp (with a slightly decreasing frequency with time). Also, the results suggested that there are certain inter-cycle cardiac variability (Fig. 4). For example, the lower-frequency component of SCG1 and SCG2 had more energy than their higher-frequency component during the first cardiac cycle. However, in the second cycle, more energy of SCG1 was concentrated in the higher-frequency component. The source of this variability is unknown but may be related to the known heart beat variability [16]. Fig. 5.b shows the SCG IF estimated by PCT and SPWVD. This figure shows that the dominant IF of SCG1 was time-dependent and significantly varied among the two cardiac cycles shown. For the first cycle, the dominant frequency fell from 21 Hz to below 10 Hz, while in the second cycle, the dominant frequency started around 20 Hz then peaked to 37.5 Hz, and finally dropped to below 10 Hz. SCG2 had a slight downward trend from 22 Hz to 19 Hz in both the first and second cardiac cycles before dropping to below 10 Hz. While these trends may be of diagnostic value, further investigations would be needed to see if they are consistent for several cardiac cycles and patients. The TFD techniques under consideration were used to estimate the dominant frequency of the SCG from 8 healthy subjects (Table IV). The data suggests that there is a broad range of IF in different subjects. The agreement between STFT and PCT appeared to be highest followed by SPWVD while WVD had the most disagreement with other methods.

IV. CONCLUSIONS

This study aimed at comparing the performance of four different TFD approaches in estimating the instantaneous frequencies of SCG signals. Methods included STFT, PCT, WVD and SPWVD. In the current study, the temporal and spectral resolution of the SPWVD and STFT was finer and coarser than other methods, respectively. The accuracy of different methods in determining the IF was tested using synthetic test signals with known TFD, and the estimated IF was compared to actual IF values. The errors in estimating IF were highest for WVD. These results may be attributed to the limitations of WVD (e.g. negative energy values and artifacts in the TFD) and suggested that the method would not be a good choice for estimating the TFD characteristics of SCG signals. The SPWVD and PCT had the lowest error in estimating IF followed by STFT. These results may be ascribed to the fine temporal and spectral resolution of PCT and SPWVD. Therefore, this study suggested that the PCT and SPWVD would be better estimators of the frequency content of SCG signals. TFD of an actual SCG was also estimated using PCT and SPWVD. Results showed that this signal had two primary frequency components that are slightly time-varying. Further studies may be warranted to investigate the frequency contents of SCG signals in health and disease.

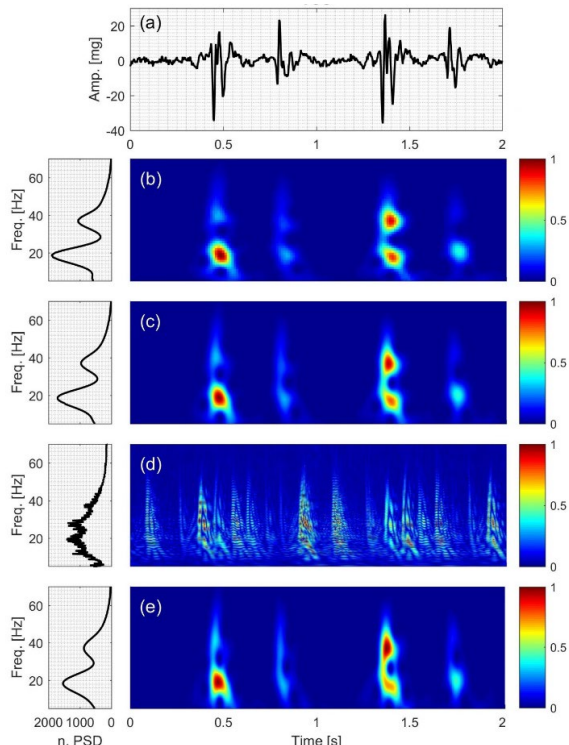

Fig. 4. Actual SCG signal: (a) Time series. Time-frequency distribution using (b) STFT, (c) PCT, (d) WVD, and (e) SPWVD, respectively.

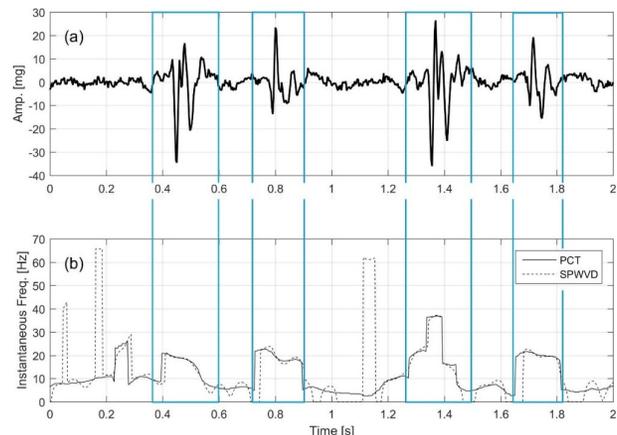

Fig. 5. (a) Time representation of SCG signal, (b) Estimating signal instantaneous frequency using PCT and SPWVD.

TABLE IV. DOMINANT FREQUENCY (HZ) OF THE ACTUAL SCG SIGNALS CALCULATED USING STFT, PCT, WVD, AND SPWVD.

| Subject # | STFT | PCT | WVD | SPWVD |
|---|---|---|---|---|
| 1 | 6.25 | 6.15 | 3.93 | 3.82 |
| 2 | 30.00 | 30.28 | 21.42 | 30.40 |
| 3 | 11.25 | 11.32 | 16.12 | 12.92 |
| 4 | 17.50 | 17.23 | 17.85 | 15.88 |
| 5 | 28.75 | 28.31 | 26.95 | 27.82 |
| 6 | 7.50 | 7.63 | 18.95 | 7.38 |
| 7 | 20.00 | 19.94 | 13.05 | 19.57 |
| 8 | 33.75 | 32.98 | 23.63 | 31.38 |

ACKNOWLEDGEMENTS

This study was supported by NIH R44HL099053.